\documentclass[a4paper]{jpconf}
\usepackage{graphicx}
\usepackage{amsmath}
\newcommand\be{\begin{equation}}
\newcommand\ee{\end{equation}}
\newcommand{\bea}{\begin{eqnarray}}
\newcommand{\eea}{\end{eqnarray}}
\newcommand{\nn}{\nonumber}
\usepackage{color}

\begin{document}
\title{Classical instability in Lovelock gravity}

\author{P. Suranyi and L.C.R. Wijewardhana}

\address{Department of Physics, University of Cincinnati, Cincinnati, Ohio 45220, USA}

\ead{suranyp@uc.edu}

\begin{abstract}
We introduce a simple method for the investigation of the classical  stability of static solutions with a horizon in Lovelock gravity. The method is applicable to the investigation of high angular momentum  instabilities, similar to those  found by Dotti and Gleiser  for Gauss-Bonnet black holes.  The method does not require the knowledge of the explicit analytic form of the black hole solution.  In this paper we apply our method to a case where the explicit solution is known and show that it identifies correctly the resulting unstable modes.\end{abstract}

\section{Introduction}
Theories formulated in more than 4 space-time dimensions have been used to provide a unified description of particle interactions.    Kaluza \cite{kaluza}  and Klein \cite{klein}, in their celebrated papers, unified  electromagnetism with gravity in  four dimensions by starting from five dimensional Einstein action and compactifying one space dimension on a circle.  Schwarzschild black hole solutions were generalized to D dimensional Einstein gravity by Tangherlini \cite{tanger}. Once it was realized that superstring theories, candidate quantum theories of gravity capable of unifying all  known forces, are consistent only in ten dimensions the study of higher dimensional models became more important and relevant. Ten dimensional Einstein action arises in the zero slope limit of superstring theory. However, perturbative contributions to string theory generate corrections to the Einstein action \cite{corrections}.  The correction terms contain higher powers of the curvature tensor.  Now most    gravity  theories with  higher order  curvature corrections have ghost excitations, with the notable counterexample, the Lovelock Lagrangian \cite{lovelock}.  The equations of motion in Lovelock gravity contain only second order derivatives and the metric is consistent with unitarity. We can write the Lovelock Lagrangian as
\be\label{lseries}
L=\frac{1}{16\pi G_D}\sum_{n=0}^\infty \alpha_n L_n.
\ee
where the terms, $L_n$, are constructed from antisymmetric combinations of powers of the curvature tensor as
\be\label{lovelockterm}
L_n=\frac{1}{2^n}\delta^{a_1b_1,a_2b_2,...,a_nb_n}_{c_1d_1,c_2d_2,...,c_nd_n}\prod_{i=1}^n {R_{a_ib_i}}^{c_id_i},
\ee
  where ${R_{ab}}^{cd}$ is the curvature tensor and the generalized Kronecker delta symbol. $\delta$, is the unique tensor that is antisymmetric under the exchange of {\it pairs}  $a_ib_i$ and $a_jb_j$ {\em or pairs} of $c_id_i$ and $c_jd_j$ and symmetric in the exchange of any $a_i$ and $ b_i$ or $c_i$ and $d_i.$  The $n=0$ term of (\ref{lseries}) corresponds to the cosmological constant, while the $n=1$ term is Einstein gravity ($\alpha_1=1,$ and $L_1=R$). The term $L_2$ is the Gauss-Bonnet term. Only a finite number of terms,  [$(D-1)/2$], contribute to the equation of motion at any fixed $D$.
  
  As a rule, actions are built based on general principles, such as symmetry and causality.  All terms satisfying these requirements and built from the appropriate fields should be included in the Lagrangian. In this sense, there is no reason, why higher order Lovelock terms should be excluded from (\ref{lseries}). As the dimension of the ratio of coupling constants $\alpha_{n+1}/\alpha_n$ is [length$^2$], higher order terms are of higher dimensions and, as a rule, their contributions become important at shorter distances. In other words, solutions in Lovelock gravity reduce to solutions in Einstein gravity when $r^{2n}>>\alpha_n$ for all $n\leq[(D-1)/2]$.
  
  Analytic spherical black hole solutions in 5 dimensional Gauss-Bonnet (GB) gravity were found by Boulware and Deser \cite{boulware}.  The surprising property of GB black holes is the existence of a minimum mass $M_{\rm min}$, proportional to the GB coupling, $\alpha_2$.  At the minimal mass the radius of horizon shrinks to zero and the Hawking temperature vanishes, making minimal mass black holes thermodynamically stable~\cite{Myers}.  However,  Gleiser and Dotti have shown \cite{dotti}, based on the decomposition of excitations by  Kodama, Ishibashi, and Seto \cite{kodama}\cite{ishibashi} that below a critical mass, $M_c>M_{\rm min}$ GB black holes have a classical instability against high harmonics excitations.  To find the location of instability in \cite{dotti} a Schrodinger-type differential equation was derived and solved for linearized excitations. The equation is diagonal in, and as usual, the  potential is dependent on the five dimensional generalized angular momentum, $L$.   Thus, $M_c$ is also dependent on $L$. Beroiz, Dotti and Gleiser \cite{beroiz}, pointed out that the onset of overall instability, which is attained at the maximum of $M_c(L)$ over all admissible $L$ values, is reached in the $L\to\infty$ limit.    Konoplya and Zhidenko \cite{konoplya} showed that there are the no other instabilities of the 5-dimensional GB black hole than the ones found by Dotti and Geiser.    In the present paper we apply a simplified method to the investigation of the classical stability of 7-dimensional spherical symmetric black holes in $n=3$ Lovelock gravity.  The exact solution for such black holes was found in \cite{dehghani}.  Takahashi and Soda generalized the formulation of Geiser and Dotti and found a master equation and general conditions for the stability of higher order Lovelock theories \cite{takahashi1}\cite{takahashi2} for positive Lovelock couplings.  Camanho, Edelstein, and Paulos completed a more general investigation of the stability of higher order Lovelock black holes \cite{camanho2}.  
  
  All the above investigations rely on finding an analytic solution to the Lovelock equations. However, other static solutions with an event horizon (dubbed `black objects' below), for which such exact solutions are not known, are also of interest. The principal aim of this paper is to demonstrate that a simple analytic method for the investigation of high harmonics instabilities can be applied and high harmonics instabilities can be found even in the absence of analytic solutions of the Lovelock equations. We will briefly discuss an example for a geometry in which such a problem arises at the end of the next section.  Using these examples, we wish to check the correctness of our conjecture that `inert' black objects do not exist even in the limiting sense. Stated more precisely, suppose that the radius of horizon of a black object of a certain topology has a range $0<r_h<\infty$.  If its AdM mass is finite and its Hawking temperature vanishes in the limit $r_h\to\,0$, than it approaches an inert state. Lovelock black holes for positive Lovelock couplings do have these properties.  Our conjecture is that for all black objects, when they exist, the inert state is always shielded by a classical high harmonics instability at a finite radius of horizon. In other words, before the object reaches its inert state by Hawking radiation it will undergo a catastrophic disintegration due to the classical instability.

  The classical stability of black holes and other black objects is not merely of academic interest. Sufficiently low scale gravity models could imply the production of black holes at future particle accelerators.  If space-time is more than four dimensional, then Lovelock terms  change the spectrum of observable black holes.  In particular, classical instabilities at nonzero mass would significantly effect the experimental signature of black hole production. \cite{phenomen}
  \section{Large angular momentum instability}
  
  We will study the metric
  \be\label{metric}
  ds^2=g_{AB}\,dx^A\,dx^B=-f(r)\,dt^2+\frac{1}{f(r)}dr^2+r^2d\Omega_{D-2}
  \ee
  representing a $D$ dimensional static black hole.
  Linearized perturbations, $g_{AB}\to g_{AB}+h_{AB}$, can be classified into scalar, vector, and tensor types \cite{kodama}\cite{takahashi1}\cite{takahashi2}. In the present paper we will concentrate on scalar perturbations.  For scalar perturbations $h_{AB}=S\, \eta_{AB}(r,t)$, where $S$ is a spherical harmonic  on $S^{D-2}$. In other words, $S$ satisfies the Laplace equation corresponding to metric $g$
 \be
\Delta_g S=-\vec L^2S=-L(L+D-3) S.
\ee
As the unperturbed metric is static, the coefficients of the equation of motion are independent of $t$ and a solution for $h_{AB}$,  having exponential dependence on time can be found. In other words, we can write
\be\label{components}
h_{AB}=e^{\omega\,t} f_{AB}(r,g) S.
\ee
Gauge transformations allow us to bring $f_{AB}$ to a simple form, whereas the non-vanishing components of $f_{AB}$ are $f_{tt}(r)$, $f_{tr}(r)$, $f_{rr}(r)$, and $f_S(r)$, where we defined $f_{ij}(r,g)=g_{ij}\,f_S(r)$  \cite{dotti}. 
If the system of linearized equations for the perturbations admits a solution with $\omega^2>0$ (runaway solution) then  the metric $g_{AB}$ is unstable. 

Geiser and Dotti derived a Schrodinger-like differential equation for a single component of $h_{AB}$ for the $D=5$ GB black hole metric.  The energy eigenvalue of the equation was $E=-\omega^2$, where $\omega$ was defined in (\ref{components}). As usual,  the Schrodinger potential depended on $\vec L^2$.  They found test functions, such that the expectation value of the Hamilton operator was negative, establishing the existence of runaway solutions. Such solution existed,  provided the ADM mass of the black hole was smaller than a critical value, $M_c >M_{\rm min}$.  At the critical mass the instability was restricted to the horizon.  If the mass was smaller the instability was extended up to a radius, outside of the horizon.  The critical mass depended  on angular momentum, taking its largest value at $L\to\infty$. 
This fact was used in a subsequent paper, \cite{beroiz}, to find a simplified method of finding high angular momentum instabilities, which only relies on the limiting behavior  of the potential at $L\to\infty$.  After the elimination of all but one component of the perturbations the equation can be brought to the form
\be\label{sch}
-f(r)\frac{d}{dr}\left[f(r) \frac{d}{dr} h(r)\right] +V h(r)=-\omega^2h(r),
\ee
where the potential has the following asymptotic behavior at large $L$ \cite{beroiz}:
\be\label{potential}
V=L^2 f(r)v_L(r)+v_0(r)+...,
\ee
Then it is easy to see that if we find an $r_0$, such that $f(r_0)v_L(r_0)<0$ then the lowest eigenvalue of (\ref{sch}) must be negative as $\omega^2\simeq O(L^2)$ for large $L$.  Then the static solution is unstable.  Near the horizon $f(r)\simeq (r-r_h)f'(r_h)$  (where $r_h$ is the radius of horizon) and  the question of stability hinges on the sign of $f'(r_h)v_L(r_h)$ .   Then the onset of the instability is determined by the largest zero of $f'(r_h)v_L(r_h)$.
 
 A significant complication presents itself when applying the above method to solve the stability problem when  the system of second order differential equations for perturbations cannot be decomposed  to yield a single second order equation.  For such a problem we propose using an alternative method of expanding the differential equations first in $L^{-2}$ and $\omega^{-2}$, keeping the leading order contributions only, and then solving the truncated equations for a single component.  This can provide an expression for $v_L(r)$, which allows one to tackle the stability problem.  We will apply this method to the investigation of $D=7$ black holes in Lovelock gravity though in this particular case we have been able to find the exact potential near the horizon. To prove our point we will avoid using the explicit form of the known solution during our investigation.  We will only use properties derived from the equations of motion, such as asymptotic expansions and horizon expansions, from which one can determine the potential $v_L$ near the horizon. 
 
As an example, it is instructive to apply our method to the study of the classical instability of {\em black branes} in Lovelock gravity.  These objects are defined as extensions of black holes into $n$ additional compact dimensions with metric
\be\label{fghmetric}
ds^2=-f(r)dt^2+\frac{g(r)}{f(r)}dr^2+r^2d\Omega_{d}+h(r)\gamma_{ij}dw^{i}dw^{j},
\ee
where $\gamma_{ij}$ is an $n$ dimensional metric for  compact coordinates,  $f(r)$ is assumed to have a zero at some $r>0$, and $h(r)$ is bounded.  Though analytic solutions for black branes are not in general known in Lovelock gravity, a numerical investigation of the static solutions is quite feasible.   The numerical investigation can provide a one-to-one correspondence between the radius of horizon and the ADM mass.  Then one can perform an analytic investigation of the instability in the infinitesimal neighborhood of the horizon because the metric functions have a unique analytic horizon expansion.  Thus, one can also determine the horizon expansion of $v_L(r)$ in arbitrary order of $r-r_h$.  This, combined with the numerically determined correspondence between the radius of horizon and the ADM mass allows one to find the onset of instability as a function of mass.  We will investigate the instability of black branes in Lovelock gravity, using this method, in a subsequent publication.  
 
 \section{Instability of $D=7$ Lovelock black holes}
 
 At $D=7$ the first four terms of series (\ref{lseries}) contribute to the equations of motion.  An $S_5$ symmetric black hole solution depends on the two constants, $\alpha_2$ and $\alpha_3$, in addition to the Newton's constant, and the ADM mass.   Static, spherically symmetric black hole solutions in 7-dimensional Lovelock gravity were found by Dehghani and Shamirzaie \cite{dehghani}.  The solutions of black holes  with third order Lovelock terms were classified by Camanho, Edelstein and Paulos \cite{camanho3}. 
 
 For $D=7$ metric (\ref{metric}) leads to the following equation of motion:
 \be\label{equationofmotion}
 {\cal{ G}}_t^t={\cal{G}}_r^r=\frac{5}{2r^5}\left\{f'(r)\,c(r)-b(r)]\right\}=0,
 \ee
 where
   \bea\label{cr}
    c(r)&=&r^4+24\, r^2\,\alpha_2[1-f(r)]+72\,\alpha_3[1-f(r)]^2,\nn\\
    b(r)&=&4\,r\,[1-f(r)][r^2+6\alpha_2.(1-f(r)).
    \eea
    
Depending on the choice of Lovelock couplings $\alpha_2$ and $\alpha_3$ there are at most three solutions of (\ref{equationofmotion})
\be\label{solution}
f(r)=\frac{1}{12\alpha_3}\left[2(6\alpha_3+r^2\alpha_2)-z\left(\frac{2}{Y}\right)^{1/3}r^4(\alpha_3-2\alpha_2^2)+z^*(2\,Y)^{1/3}\right],
\ee
where
\be
Y=W+\sqrt{W^2+2r^{12}(\alpha_3-2\alpha_2^2)},
\ee
\be
W=r^6\alpha_2\left(4\alpha_2^2-3\alpha_3\right)-18\alpha_3^2m,
\ee
and the three possible solutions are defined by the choices $z=1, e^{2\,i\,\pi/3}, e^{-2\,i\,\pi/3}$. 

$m$ is the coefficient of the asymptotic expansion series
\be\label{ass}
f(r)\simeq 1+a \,r^2 -\frac{m}{r^4}+\frac{b}{r^{10}}+...
\ee
$m$ is related to the ADM mass, $M_{\rm ADM}$, as
 \be\label{admmass}
m=M_{\rm ADM} \frac{16\pi G_7}{5\Omega_5}.
 \ee
 As $m$ is proportional to $M_{\rm ADM}$ we will refer below to the former as ''mass'', as well.
 The solution is asymptotically Minkowski if $a=0$ in (\ref{ass}), while it is AdS (dS) type if $a$ is positive (negative).
Then the radius of horizon (when there is a horizon exists) is 
\be\label{radius}
 r_h=\sqrt{- 6\alpha_2+\sqrt{36\alpha_2^2-24\alpha_3+m}}, 
 \ee

Now we turn to the analysis of the stability of  the $D=7$ black hole in Lovelock gravity, using our simplified method.  As we indicated above, we will refrain from using the explicit form of the black hole solution, given by (\ref{solution})-(\ref{radius}) but rather rely only on (\ref{equationofmotion}) and expansions based on it. These are readily available even for metrics for which the equations cannot be solved.  Let us first find the coefficients of the asymptotic series expansion in $r$. It is easy to see, by replacing $f(r)$ in  (\ref{equationofmotion}) by an asymptotic series in powers of $r^{-2}$, that the only possible leading order nonzero coefficients are those given in (\ref{ass}).  Furthermore, we find that the only possible solutions for coefficient $a$ are
\be\label{acases}
a=\begin{cases} 0, &\text{ asymptotically Minkowski solution,}\\
\frac{1}{4\alpha_3}\left(\alpha_2-\sqrt{\alpha_2{}^2-\frac{2}{3}\alpha_3}\right),&\text{asymptotically AdS or dS solution,}\\
\frac{1}{4\alpha_3}\left(\alpha_2+\sqrt{\alpha_2{}^2-\frac{2}{3}\alpha_3}\right),&\text{asymptotically AdS or dS solution.}
\end{cases}
\ee
The constant $m$ (the ''mass'') is undetermined, but once it is fixed it determines all further asymptotic expansion coefficients for a given choice of $a$ (as long as that value is admissible for the given Lovelock couplings).  Assuming a finite radius of convergence, and no naked singularity (for real $r$), $a$ and $m$ determine the solution uniquely.  We will find that in a similar way the horizon expansion is also defined uniquely for given radius of horizon, $r_h$, and Lovelock couplings.  (\ref{acases}) shows that besides a solution with Minkowski asymptotics, which exists for all Lovelock couplings,   one solutions with AdS and  one with dS asymptotics exist if $\alpha_3<0$.  Furthermore, if $0\,<\,\alpha_3\, <3\,\alpha_2{}^2\,/\,2$ then two AdS solutions exist if $\alpha_2\,>\,0$ and two dS solutions if $\alpha_2\,<\,0$. This result agrees with the classification of \cite{camanho4}.

The horizon expansion of $f(r)$ can also be found from (\ref{equationofmotion}), without the knowledge of an analytic solution.  In particular, 
 one obtains
 \bea\label{horizon1}
 f'(r_h)&=&\frac{b(r_h)}{c(r_h)}=\frac{4r_h(r_h^2+6\alpha_2)}{r_h^4+24r_h^2\alpha_2+72\alpha_3},\nonumber\\
 f''(r_h)&=&\frac{b'(r_h)c(r_h)-c'(r_h)b(r_h)}{[c(r_h)]^2}=-4\frac{1}{(r_h^4+24r_h^2\alpha_2+72\alpha_3)^3}\big[5r_h^{10}+90r_h^8\alpha_2\nn\\&+&432r_h^6(2\alpha_2^2-\alpha_3)-15552 r_h^2\alpha_3^2-31104\alpha_2\alpha_3^2+864r_h^4\alpha_2(8\alpha_2^2-9\alpha_3)\big].
 \eea
 Higher order expansion terms can be readily obtained in a  similar manner. The horizon expansion is unique for given Lovelock couplings and a given choice of $r_h$.  Again, if there is no naked singularity at or above the horizon the series has a finite radius of convergence and can be continued analytically to all positive $r>r_h$.  Consequently, it will coincide with  the analytic continuation of one of the asymptotic solutions.  Then, by necessity, there will be a one-to-one map of admissible values for $r_h$ and of mass parameters, $m>0$.  This map cannot be determined analytically, without the knowledge of the exact solution, but can be obtained by numerical integration of the equation(s) of motion.
 
 As is seen from (\ref{acases}) if $\alpha_3\, <3\,\alpha_2{}^2$ then there are three asymptotic solutions of (\ref{equationofmotion}) . However, for every choice of $r_h$ there is an unique horizon expansion.  Then two of the three asymptotic solutions do not have horizons. They must end in naked singularities at some finite $r_0\geq0$.  In case the equation of motion cannot be solved analytically numerical integration starting from the horizon selects which of the  three asymptotic behaviors (\ref{acases}) corresponds to the solution with a horizon.
 
 A further comment, concerning the horizon expansion is that the expansion diverges at 
 \be\label{crh}
c_0(r_h)= r_h^4+24r_h^2\alpha_2+72\alpha_3=0.
 \ee
 The largest root of this equation for $r_h{}^2>0$ is
 \be\label{constraint1}
 r_s{}^2=-12\alpha_2+\sqrt{144\alpha_2{}^2-72\,\alpha_3}.
 \ee
 We define $c_0(r)=c(r)|_{f(r)=0}$.  As we will see below, for a given choice of the Lovelock couplings  $r_s$ provides a lower or upper bound on physical values of $r_h$. 
   We prove now the following (only to cases when the right hand side of (\ref{constraint1}) is real): If  at all real $r>r_h$ the solution is regular then $c(r)$ does not change sign.  The proof is very simple: If there is no singularity above the horizon then $f'(r)$ is bounded at any finite $r>r_h$,  Then it follows from  (\ref{equationofmotion}) that $c(r)=0$ implies $b(r)=0$, as well. The only simultaneous solution of these two equation for $r^2$ and $f(r)$ is $r^2=0$ and $f(r)=1$. 
 
 In particular $c(r_h)>0$ for a solution with Minkowski asymptotics, because then $f(r)\to1$ and  $c(r)\simeq r^4>0$  when $r\to\infty$. Similarly, one can see that if $\alpha_3<0$ any regular solutions with dS or AdS asymptotics must have $c(r)<0$.  In particular, $c(r_h)<0$.  However, as shown by (\ref{crh}), for large $r_h$ $c(r_h)>0$. Thus, at $\alpha_3<0$ for asymptotically dS or AdS solutions the radius of horizon of a black hole is limited to $r_h<r_s$. For Minkowski asymptotics the reverse is true, $r_s>r_h$.

 Returning to (\ref{constraint1}), it follows that the $r_h\to 0$ limit is possible only if $\alpha_3 > 2\alpha_2{}^2$ (Minkowski asymptotics only) or $0<\alpha_3 <2\alpha_2{}^2$ and $\alpha_2>0$ (Minkowski or AdS asymptotics). Since our main interest is the fate of black holes (or black branes) in the limit of $r_h\to 0$ we will concentrate on solutions at $\alpha_3>0$, though instabilities can also be analyzed in a similar way when $\alpha_3<0$ . 
  
 Writing the equation of motion, linearized in time dependent perturbations as
 \be
 {\cal G}_A^B\to {\cal G}_A^B+ {\cal H}_A^B ,
 \ee
 the equations for the perturbations take the general form
 \be
 {\cal H}_A^B=\sum_{A,B}\left[ \gamma_{r,L,\omega}^0\,f_{AB}+\gamma_{r,L,\omega}^1\,f_{AB}'+\gamma_{r,L,\omega}^0\,f_{AB}''\right]=0,
 \ee
 where the coefficients $ \gamma_{r,L,\omega}^k$, $k=0,1,2$, depend on the metric components of the static solution.
Rather than solving this system of equations exactly we truncate these equations, by expanding $\gamma_{r,L,\omega}^k$ in
$1\,/\,\vec L^2$ and in $1\,/\,\omega^2$ and keeping only the leading order term of the expansion. The truncated equations have the form 
 \bea\label{perteq}
 {\cal H}_t^t&=&-\frac{\vec L^2}{2\, r^7}[r\,f(r)\,c(r)\,f_{rr}+c'(r)\,f_S]+\frac{5\,c(r)\,f(r)}{2\,r^6}[f_S''-r\,f(r)\,f_{rr}']\simeq\,0
,\nonumber\\
  {\cal H}_r^t&=&-\frac{\vec L^2}{2\, r^6\omega}c(r)\,f_{rt}-\frac{5\,c(r)\,f(r)}{2\,r^6}f_S'\simeq\,0,\nonumber\\
   {\cal H}_r^r&=&-\frac{c(r)}{2\, r^6f(r)}[5 \,\omega^2 f_S-10\,\omega^2\,r\,f(r)\,f_{rt}-{\vec L}^2f_{tt}]-\frac{c'(r)\,\vec L^2}{2 \,r^7}f_S\nn\\&-&\frac{5}{4\,r^6}\{2\,r\,c(r)\,f_{tt}'-[2c'(r)f(r)+c(r)f'(r)]\,f_S'\}\simeq\,0,\nonumber\\
    {\cal H}_\psi^\theta&=&\frac{f(r)\,c'(r)}{8\, r^5}f_{rr}-\frac{c'(r)}{8 \,r^5f(r)}f_{tt}+\frac{c''(r)}{8 \,r^6}f_S\simeq\,0,
    \eea
    where $c(r)$ was defined in (\ref{cr}).
  
        The rest of the equations, ${\cal H}_A^B=0$, are not independent from the above four. We present the equations without truncation in the Appendix.  
        
        Now taking the combinations of equations (\ref{perteq}) 
        \be
            {\cal H}_{0}=\frac{2\,r^6\,f(r)}{5 c(r)}[ {\cal H}_t^t-\frac{10\,r\,\vec L^2}{\omega^2\,f(r)}  {\cal H}_r^t+ {\cal H}_r^r+\frac{4\,c(r)}{r\,c'(r)}\vec L^2  {\cal H}_\psi^\theta]\simeq\,0
            \ee
we obtain the following second order differential equation for $f_S$.
   \be\label{scheq1}
 -f(r)\frac{d}{dr}\left[f(r) \frac{d}{dr} f_S\right]+\vec L^2 f(r) v_L(r) f_S=-\omega^2f_S,
   \ee
   where  
  \be\label{diffeq}
   v_L(r)=\frac{2\,[c'(r)]^2-c(r)c''(r)}{5\,r\,c(r)\,c'(r)}.
   \ee
 Note that (\ref{diffeq}) coincides with the eq. (48) of \cite{takahashi1} derived using the explicit solution of the Lovelock equations. 
   A term, proportional to $f_S'$, and finite in the limit $L, \omega\to\infty$, has been omitted from (\ref{diffeq}) as it can be eliminated by factoring out a multiplier from $f_S$. This would only contribute by a sub-leading, $O(1)$, term to the potential. 
   
      Note that our method enabled us to derive the functional form of $v_L(r)$ without assuming the exact functional form of the Lovelock black hole solution. Furthermore, dropping  the $O(1)$ derivative terms from (\ref{perteq}),  one obtains four algebraic equations for the perturbations $f_{AB}$, of which ${\cal H}_t^t=0$ and $  {\cal H}_r^t=0$ imply that in leading order the perturbation $f_{rt}$ vanishes. Then dropping $f_{rt}$, along with other non-leading order contributions from equations $ {\cal H}_r^r=0$ and ${\cal H}_\psi^\theta=0$ we obtain a system of two {\em linear algebraic equations} for $f_{tt}$ and $f_S$.  Eliminating $f_{tt}$ leads to the single equation for $f_S$ which reads as
   \be\label{instabeq}
   \vec L^2 f(r) v_L(r) f_S=-\omega^2f_S,
   \ee
with $v_L(r)$ given in (\ref{diffeq}), with $c(r)$ replaced by $c_0(r)$.  

To summarize, our procedure consists of the following steps: (1) Derive the system of equations for linear perturbations form the Lovelock equations (see the appendix for the complete equations), (1) Drop all derivative terms from the system, (2) Expand the coefficients of perturbations $f_{AB}(r)$ to leading order in $L^{-2}$ and $\omega^{-2}$, (3) Discard equations in which the only leading order terms contain off diagonal perturbations, and omit the terms containing off diagonal perturbations from the rest of the equations, (4) Calculate the determinant, $Det$, for a subset of independent linear equations for the diagonal perturbations $f_{AA}(r)$, (5) Solve  the equation $Det=0$ for $\omega^2$, to obtain an equation for the potential
\be
\omega^2= -f(r)L^2v_L(r).
\ee
  
  Equation (\ref{instabeq}) verifies that our simplified procedure for finding $v_L$ provides the same leading order potential as \cite{takahashi1},\cite{takahashi2}). While this method of finding the potential does not constitute a significant simplification compared to the ''exact''  derivation of (\ref{scheq1}), it is the only way to derive the leading order potential, $v_L(r)$ for a more complicated metric, such as (\ref{fghmetric}). Note that for such a metric, due to the larger number of perturbations, $v_L(r)$ is not unique, but the equation $Det=0$ has several solutions $v_L^i(r)$, each leading to independent linear perturbations.  If any of $f(r)v_L^i(r)<0$ at some $r=r_0$, then the static solution is unstable.  We will apply this method to the study of instabilities for black branes in a subsequent publication. 
   
 Now consider the  case when the exact static solution, $f(r)$,  is not known.  Then, finding the the leading order potential, $v_L(r)$, as a function of the metric components does not yet allow us to find instabilities.  However, we can calculate $v_L(r)$ in the neighborhood of the horizon by making use of the horizon expansion (\ref{horizon1}).    $v_L(r)$ can  readily be obtained from (\ref{diffeq}) if we substitute
        \bea\label{cder}
        c_0(r)&\to&r^4+24\,\alpha_2\,r^2+72\,\alpha_3,\nn\\
        c_0'(r)&\to&4\big[r^3-6\,f'(r)\,(\alpha_2r^2-6\,\alpha_3)\big]\nn\\
        c_0''(r)&\to&12\big\{r^2+4\,\alpha_2-8\,r\,\alpha_2\,f'(r)+12\,\alpha_3[f'(r)]^2-2\,f''(r)[r^2\alpha_2+6\,\alpha_3]\big\},
        \eea
        where, for simplicity, we dropped the subscript $h$ from $r_h$. We can use the expressions $f'(r_h)$ and $f''(r_h)$ in (\ref{cder}) as they are given in (\ref{horizon1}). In fact, as shown by (\ref{instabeq}), the solution is unstable near the horizon if the quantity $f(r)v_L(r)\simeq (r-r_h)f'(r_h)v_L(r_h)<0$. Since $r-r_h>0$ the solution is unstable in the immediate neighborhood of the horizon if $V_L(r_h)=f'(r_h)v_L(r_h)<0$. Consequently, from here on we will calculate the "effective potential", $V_L(r_h)$, rather than $v_L(r_h)$.
    
For $\alpha_2\not=0$ we obtain
\bea\label{pmpot}
V_L(r)&=&\frac{R\pm6}{5\sqrt{R} (R^2\pm12\, 
    R+144-72\beta) (R^2\pm24\, R + 72\, \beta)^3}\Big\{5 R^{6} + 72\, 
  R^4 (6 - 43\beta )\nn\\& &+ 5184\, 
 R^2 (32 - 72\beta+3\beta^2-
   1119744 (2 - \beta)\beta^2
 \pm\big[1728\,R^3 (16 - 33\beta)\nn\\&-&  62208\,\beta \, 
  R (16 - 3\beta)- 60 R^{5} ]\Big\},
    \eea
    where we introduced the dimensionless variables $R=r_h{}^2/\,|\alpha_2|$ and $\beta=\alpha_3/\,\alpha_2{}^2$, and where  $\pm1={\rm Sign}(\alpha_2)$.   We will discuss the potential at $\alpha_2=0$, which is much simpler, later. 
   Then, as we pointed out in the previous section, a sufficient condition for the existence of high harmonics scalar instability at any given value of $\alpha_2$ and $\alpha_3$ is the existence of $R_c\geq R_{min}$, where $R_{min}$ is the minimum value of the radius of horizon, associated with the minimum mass, such that $V_L(R_c)<0$. We will concentrate on the stability of the solution near the horizon. Note that inert black holes would have $R_{min}=0$.
   
   The asymptotic expansion of $V_L(R)$,
   \be\label{pot2}
   V_L(R)\simeq \frac{4}{R^{3/2}}>0
   \ee
   shows that the potential is positive at large values of $R$.  That, of course, follows from the stability of  the Schwarzschild-Tangherlini black hole, approached by any Lovelock black hole when $R\to\infty$.  To find instabilities one needs to investigate positive, 0, and negative values of $\alpha_2$ separately. 

   As we have shown earlier, when $\alpha_2>0$ and $\alpha_3>0$  (\ref{constraint1}) does not forbid the radius of horizon, $R$, to approach zero. However, there is always a single positive zero, $R_c>0$, such that $V_L(R_c)=0$ and the solution  becomes classically unstable at $R<R_c$. $R_c$ is given as the solution of a 6th order equation.  Furthermore, $V_L(R)$ also has a pole at $0<R_p<R_c$, $R_p=6(\sqrt{2\beta-3}-1)$, if $\beta>2$. The pole corresponds to the zero of $c_0'(r)$ in (\ref{diffeq}). Then it also has a second zero at $R_2<R_p$.  For small $R$ it is always negative (provided $\beta>0$)
   \be
   V_L(R)\simeq -\frac{1}{5\sqrt{R}\alpha_3}+..<0
   \ee
   Then for $\alpha_3>2$ the potential is negative at $R<R_2$ and at $R_p<R<R_c$.  Though the potential is positive at the horizon when $R_2<R<R_p$ one can show with further calculations that it has a range of instability above the horizon. Thus, the static solution is unstable at $R<R_c$. Whether or not $R_{min}<R_c$ there cannot be any inert black hole for $\alpha>0$ and $\alpha_2>0$.  Of course, the exact solution shows that for this range of the Lovelock couplings $R_{min}=0$. The same thing can be discovered
 in a more general case, e.g. for a metric like (\ref{fghmetric})  by connecting the radius of horizon with $m$ by numerical integration.  
 
   In Fig.1, the potential $V_L(R)$ is plotted as a function of $R=r_h{}^2\,/\,|\alpha_2|$ for $\alpha_2>0$ and two typical  values of $\beta= \alpha_3\,/\,\alpha_2{}^2$, $\beta=1$ (blue curve), at which value the potential has a single zero and approaches $-\infty$ as $R\to 0$, and $\beta=30$ (red curve).  At the latter value of $\beta$, besides the zero at $R=R_c$ there is a pole and a second zero of the potential. The location of the pole is shown by the vertical red line.  The solution is unstable in a neighborhood of the horizon for all $\alpha_3>0$. This has already been pointed out  in \cite{takahashi1}.
  
   \begin{figure}
\begin{center}
\includegraphics{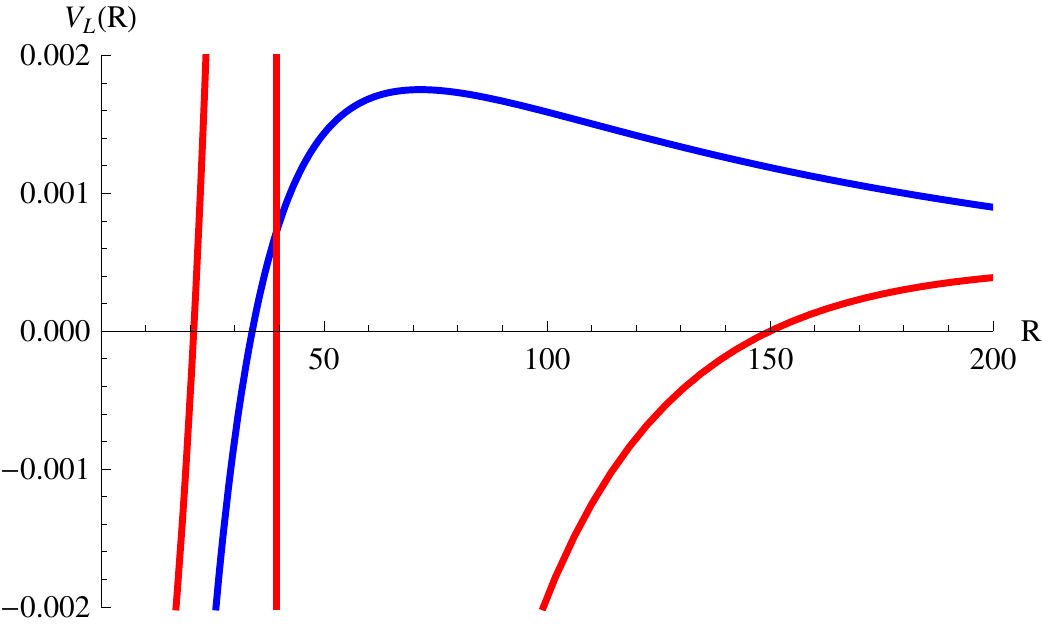}
\end{center}
\caption{\label{vlplus1}$V_L(R)$ as a function of $R=r_h{}^2\,/\,|\alpha_2|$ for $\alpha_2>0$ at two typical values of $\beta=\alpha_3/\alpha_2{}^2$, $\beta=1$ (blue curve) and $\beta=30$ (red curve).  The red vertical line signifies the location of the pole of $V_L(R)$,  $R_p$.}
\end{figure}

   For $\alpha_2=0$ and $\alpha_3>0$ we obtain
    \be
    V_L^0=\frac{4\,r[5\,r^{12}-3096\,r^8\alpha_3+15,552\,r^4\alpha_3{}^2+1,119,744\,\alpha_3{}^3]}{5\,(r^4-72\,\alpha_3)(r^4+72\,\alpha_3)^3},
    \ee
    where, for simplicity, as before,  we dropped the subscript $h$ from $r_h$. 
   We can introduce the dimensionless variables $\rho=r_h/(\alpha_3)^{1/4}$ and $\mu=m/(\alpha_3)^{1/2}$. Then the minimum of the mass is at $\mu_{min}=24^{1/2}\simeq 4.90$ (which value could be found numerically, had we not known the analytic solution) attained at $\rho=0$ and the potential at the horizon simplifies to
   \be
  V_L(\rho)=\frac{4\,\rho^{1/4}[5\,\rho^{3}-3096\,\rho^2+15,552\,\rho+1,119,744]}{5\,(\rho-72)(\rho+72)^3}. 
   \ee
   The potential is positive at large $\rho$ and has a zero at 
   \be
   \rho_c=\frac{1032}{5}+\frac{96}{5}\cos\left[\frac{1}{3}\tan^{-1}\left(\frac{45\sqrt{10}}{859}\right)\right]\simeq4.9769
   \ee
   where it turns negative.  When we further decrease $\rho$ the sign of the potential changes to positive again at the pole $\rho=72$ ($r_p=(72)^{1/4}\simeq2.9129$) and stays positive down to the minimum value of the horizon, $\rho=0$. In other words, the solution is unstable at the horizon in the range $\rho_p<\rho<\rho_c$.  Again, one can show that at $\rho<\rho_p$  there is a region of instability at non-vanishing radial distances (above the horizon).

 If $\alpha_2\,<\,0$ we consider only $\alpha_3>2 \alpha_2^2$.   We use the same dimensionless quantities, $\beta=\alpha_3/\alpha_2{}^2$  and $R=r_h^2\,/\,|\alpha_2|$ as for the discussion of the $\alpha_2>0$ case. 
The exact solution suggests that minimum value of $m$ is 
\be
m_{\rm min}=24 \alpha_3-36 \alpha_2^2,
\ee
and $R=R_{min}=6$ at the minimum mass. Of course, without the exact solution one can only find $R_{min}$ from numerical integration of (\ref{equationofmotion}). 
The effective potential (\ref{radius}) has a single pole at 
\be
R^2\,-12\, 
    R+144-72\beta=0
    \ee
    The positive root of this equation is $R_p=6(1+\sqrt{2\beta-3})>R_{min}=6$. The effective potential changes sign at this point.  In other words, the solution becomes unstable above the horizon. In fact, the solution becomes unstable at an even larger value of $R$, because the numerator of $V_L(R)$, (\ref{pmpot}) has a zero above the pole for all $\beta>2$.  
    \begin{figure}
\begin{center}
\includegraphics{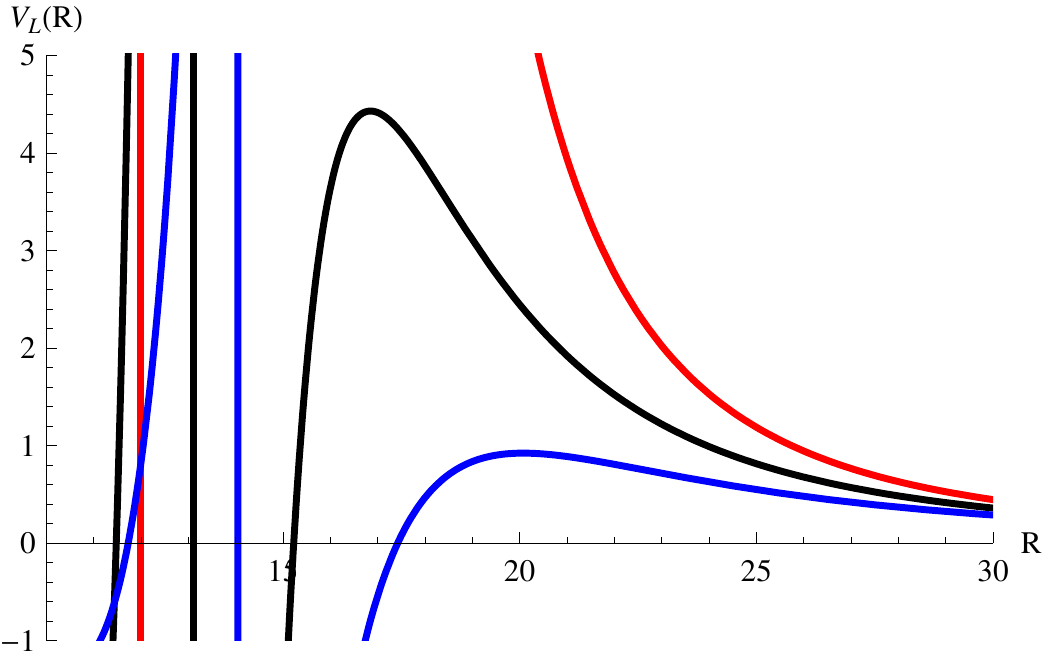}
\end{center}
\caption{\label{vlminus1}The 'effective potential,' $V_L(R)$ as a function $R=r_h{}^2\,/\,|\alpha_2|$   at the critical $\beta=2$  (red), $\beta=2.2$ (black), and $\beta=2.4$ (blue).  The location of the pole of $V_L(R)$, $R_p$, appears as a vertical line. At each $\beta>2$ the potential has a zero at $R_c>R_p$.}
\end{figure}

 Fig. 2 shows the effective potential $V_L(R)$ as a function of $R$ at three different choices of $\beta$, the "critical" value $\beta=2$ and $\beta=2.2.\,2.4$. In each case the potential turns negative at some $R>6$.  In other words there is an instability above the horizon.  If the black hole decays by Hawking radiation its radius will never reach the minimal value $R_{min}=6$.  In the critical case the potential has no zero but has a pole at  $R=12$.  In the plot the vertical lines show the location of the pole at different values of $\beta$. 
Summarizing  our results for $\alpha_3>0$ we found that in all physical cases there is a high harmonics   instability at a critical mass that is larger than the minimal mass, whether or not the minimum radius of horizon can reach $r_h=0$.

Finally we briefly discuss the case $\alpha_3<0$.  There are  three solutions of the equations of motion. One is asymptotically Minkowski while the other two are asymptotically  AdS and dS (see (\ref{acases})\,). At a given $r_h$ only one of them can have a horizon. Now $c_0'(r_h)$ does not have a zero at any $r_h>0$, The potential at the horizon, $V_L(R)$, where $R=r_h{}^2\,/\,|\alpha_2|$, is positive at $R>R_s$  turns negative at  $R=R_s$ where $c(r_s)$=0 and $r_s$ is given in (\ref{constraint1}).   $V_L(R)$ has a triple pole at that point.  Consider first the case of Minkowski asymptotics.  We proved  in Sec. 3. that $r_h>r_s$. $V_L(R)$ is positive for $R>R_s$ and the solution is stable. As the radius horizon cannot approach zero, no inert black holes exist.  Consider now case of dS or AdS. Then we must have $r_h<r_s$.  When $r_h\to 0$
\be
V_L(R)\simeq- \frac{\alpha_2}{5r\alpha_3}.
\ee
Thus at $r_h\to 0$ the solution is unstable at the horizon for $\alpha_2<0$, but stable for $\alpha_2>0$. In fact, in the latter case the solution is stable everywhere at small $r_h$. However, the minimum radius of horizon is greater than zero. This is obvious from the exact solution,  (\ref{radius}), and can be found out by integrating the equation of motion numerically. 
\section{Summary}
	We have introduced a simple method for  the investigations of  high harmonics scalar instabilities of static solutions in Lovelock gravity.  The method is based on the truncation of the system of equations for linear perturbations, keeping only terms of leading order in the generalized angular momentum.  This leads to a system of algebraic equations for the perturbations. Solving this system provides the leading order contribution to the potential, $v_L$, in the Schrodinger-type equation satisfied by perturbations.  $v_L$ can be calculated in the neighborhood of the horizon, $r_h$, without having an exact static solution, by applying a horizon expansion to the equations of motion. The terms of the horizon expansion, and consequently the leading order term of the potential,  depend only on the radius of horizon and the Lovelock couplings. As is known from the investigation of Dotti and Gaiser \cite{dotti}\cite{beroiz} that if $v_L$ turns negative for some $r>r_h$ then the static solution becomes unstable. The advantage of our method is that it can be equally applied to spaces with compact coordinates, in which static solutions are not generally known in Lovelock gravity. An example for a metric in such spaces is given in (\ref{fghmetric}). 
	
	 To demonstrate the efficiency of our method we investigated the   instability of static $D$=7 Lovelock black holes.   Since the exact black hole solution can be found \cite{dehghani},  the results for the instability of our method can be verified.    In $D=7$ the second and third order Lovelock terms contribute to the equation of motion.  Both static solutions and perturbations depend on the two Lovelock couplings, $\alpha_2$ and $\alpha_3$.  Using the horizon expansion one finds  that for every choice of the horizon radius the solution is unique.  However, from asymptotic expansions one finds a unique solution for $\alpha_3>3 \alpha_2{}^2/2$ only. For $\alpha_3<3\alpha_2{}^2$ there are three asymptotic solutions, whereas only one of them has  Minkowski asymptotics, the others are asymptotically AdS, or dS (see (\ref{acases})\,).   However, due to the uniqueness of the horizon expansion only one of these solutions can have a horizon, the others must end at low $r$ in naked singularities.  These results are in agreement with the analysis of \cite{camanho2}.

Note that our method allows one to find the instabilities as a function of the radius of horizon.  To find the onset of instabilities as a function of the ADM mass, in cases when exact solutions are not known, one needs to integrate the equation of motion to find the unique relationship $r_h=F(M_{ADM})$ numerically. 

We found that for $\alpha_3\,>\,0$ the solution always becomes unstable at some $M_c > M_{\rm min}$, where $M_{min}$ is the minimum mass at the given values of $\alpha_3$ and $\alpha_2$.  Since the critical mass is determined by a high order algebraic equation, we determined the value of the critical mass numerically.  At $\alpha_3\,<\,0$  only a solution with a dS asymptotics has a horizon.  The minimum radius of horizon is larger than zero.  There is no high angular momentum scalar instability.  However, the black hole with the minimal mass has a nonzero Hawking temperature and is consequently thermodynamically unstable.  

{\em We conjecture} that in Lovelock gravity, for all static solutions with a horizon (1) either the minimum radius of horizon, associated with nonzero minimum mass,  ($M_{min}>0$) is zero, in which case there is a critical mass, $M_c$, such that all solutions $M_{min}<M_{ADM}<M_{c}$ have classical instabilities, or (2) the minimum radius of horizon is not zero and the static solution with $M_{ADM}\to M_{min}$  is thermodynamically unstable.  In the latter case the limiting state with the  $M_{ADM}= M_{min}$ must have a naked singularity. Even in case (1) a state, like the 5-dimensional Gauss-Bonnet black hole, with $r_h=0$ would represent a naked singularity, but had the instability not existed, this state would have been reached by Hawking radiation only after an infinite time. In either case the existence of an inert massive object (with zero Hawking temperature) is dynamically forbidden. 

The method we use to find high angular momentum instabilities can be applied, with some modifications, to more complicated geometries in which one cannot even find the analytic form of the static solution, let alone solving the differential equation for linear perturbations, such as black branes in Lovelock gravity \cite{Kobayashi} \cite{ours}\cite{others}.  We intend to discuss the high angular momentum scalar instability of black branes (\ref{fghmetric}) in Lovelock gravity in a future publication.  Such an investigation may have a phenomenological interest, as well, because if in TeV scale gravity the radius of the horizon of a black hole is in some dimensions larger than the  compactification radius, then the black hole appears as a black brane. 
\ack
This work is supported in part by the DOE grant \# DE-FG02-84ER40153..  L.C.R.W. thanks  CP3 Origins Institute  in Odense  
and Aspen Center for Physics for their hospitality.  We thank Drs. Jiro Soda and Jose Edelstein for calling our attention to their results of Ref. \cite{takahashi1},\cite{takahashi2},\cite{camanho2},\cite{camanho3},\cite{camanho4} and for discussions.

   \section*{Appendix}
The complete contribution of linear perturbations to the equations of motion (with overall angle dependent factors removed) are as follows:
\bea
{\cal H}_t^t&=&\frac{f(r)c(r)}{2\,r^6}\left\{\left[-\vec L^2-5\,r \,f(r)\frac{c'(r)}{c(r)}-10\,r\,f'(r)\right]f_{rr}+\frac{5}{2\,r}\left[2\,r\frac{c'(r)}{c(r)}+r\frac{f'(r)}{f(r)}-4\right]f_{S}'\right.\nonumber\\
&+&\frac{1}{r^2}\left[5\,r\frac{c'(r)[1-2f(r)]}{c(r)f(r)}-5\frac{f'(r)}{f(r)}+2-r\frac{c'(r)}{f(r)\,c(r)}\vec L^2\right]f_S
-5\,r\,f(r)f_{rr}'+5f_S''\Bigg\}.
\eea
\be
{\cal H}_t^r=\frac{5f(r)\,c(r)}{2\,r^7}\left\{\frac{r}{5}\vec L^2\,f_{tr}+r^2f(r)\,f_{rr}+\left[1+\frac{r\,f'(r)}{2\,f(r)}\right]f_{S}-r\,f_S'\right\}.
\ee
\bea
{\cal H}_r^r&=&\frac{5\,c(r)}{2\,f(r)\,r^6}\left\{2\,r\,f(r)\,\omega^2\,f_{tr}+\left[\frac{\vec L^2}{5}+r\,f'(r)\right]f_{tt}-r\,f^3(r)\left[\frac{c'(r)}{c(r)}+\frac{f'(r)}{f(r)}\right]f_{rr}-r\,f(r)\,f_{tt}'\right.\nonumber\\
&+&\left[-\frac{c'(r)\,f(r)}{5\,r\,c(r)}\vec L^2-\omega^2+f(r)\frac{c'(r)[1-2\,f(r)]}{r\,c(r)}-\frac{f'(r)}{r}\right]f_S\nn\\&+&\left[\frac{f^2(r)\,c'(r)}{c(r)}+\frac{f(r)\,f'(r)}{2}\right]f_S'\Bigg\}.
\eea
\be
{\cal H}_t^\psi=\frac{f(r)\,c(r)}{2\,r^7}\left\{-r\,f_{rr}-\frac{c'(r)}{f(r)\,c(r)}f_S+\left[\frac{r\,c'(r)}{c(r)}+\frac{r\,f'(r)}{f(r)}-1\right]f_{tr}+r\,f_{tr}'\right\}.
\ee
\bea
{\cal H}_r^\psi&=&\frac{c(r)}{2\,r^6\,f(r)}\left\{-\omega^2f_{tr}-\left[\frac{1}{r}+\frac{f'(r)}{2\,f(r)}\right]f_{tt}+\frac{2\,f(r)\,c'(r)}{r^2\,c(r)}f_S+\left[\frac{c'(r)}{c(r)}+\frac{f'(r)}{2\,f(r)}\right]f_{rr}\right.\nn\\ &-&\left.\frac{f(r)\,c'(r)}{r\,c(r)}f_S'+f_{tt}'\right\}.
\eea
\be
{\cal H}_\psi^\theta=\frac{1}{8\,r^6\,f(r)}\left[r\,f^2(r)\,c'(r)\,f_{rr}-r\,c'(r)\,f_{tt}+f(r)\,c''(r)\,f_S\right].
\ee

\end{document}